# The study of structure, electronic and optical properties of double-walled carbon nanotube bundle under hydrostatic pressure

Xiaoping Yang[1,*] and Gang Wu[2]

[1]*National Laboratory of Solid State Microstructures and Department of Physics,*
*Nanjing University, Nanjing 210093, P. R. China*
[2]*Department of Physics and Astronomy,*
*California State University Northridge,*
*Northridge, California 91330, USA*

## Abstract

Combining a classical force field, a tight-binding model, and first-principles calculations, we have studied structural, electronic, and optical properties of double-walled carbon nanotube (DWNT) bundles under hydrostatic pressure. We find that the outer tube acts as a protection shield for the inner tube and the inner tube increases the structure stability and the ability to resist the pressure of the outer tube. Moreover, the collapsed structures of the double-walled carbon nanotube bundle called "parallel" and "in-between" are more stable than the one called "herringbone". The structural phase transition induces a pseudogap along symmetry line $\Gamma X$. Furthermore, the optical properties change greatly after the collapse and a strong anisotropy appears in the collapsed structure. This provides an efficient experimental way to detect structural phase transitions in DWNT bundles.

*Electronic address: bunnyxp@hotmail.com; Also at: Department of Physics, Huainan Normal University, Huainan, Anhui 232001, P. R. China

# I. Introduction

It is well known that the physical properties of carbon nanotubes (CNTs) strongly depend on their geometrical structure, which can easily be changed by applying pressure or strain. For example, an uniaxial strain on a single-walled carbon nanotube (SWNT) can cause a metal-semiconductor transition [1]. This behavior could be used to fabricate nanoscale electromechanical coupling devices or transducers.

Recently, the effect of hydrostatic pressure on CNT bundles has attracted much attention in experimental [2–10] and theoretical material science [11–15]. Studies of the SWNT bundles indicate that with increasing hydrostatic pressure Raman peaks shift to higher frequencies and the radial breathing mode (RBM) disappears from the spectrum above the critical pressure, showing the structural phase transition (SPT). On the other hand, there are only a few studies on the effect of hydrostatic pressure on double-walled carbon nanotube (DWNT) bundles, which are the simplest multi-walled carbon nanotubes (MWNTs). Very recently, Raman measurements on DWNTs under hydrostatic pressure [2–4] have found that the RBM intensity of the outer tubes decreases rapidly with increasing pressure, exhibiting a similar behavior to that of SWNT [5–13]. The inner tubes appear to be considerably less sensitive to pressure, *e.g.*, the pressure coefficient of the inner tube is 45% smaller than that of the outer tube.

Pure and highly crystalline DWNTs have been synthesized by M. Endo *et al.* [16]. By using Raman spectral analysis they find two sets of DWNT pairs having inner-to-outer diameter (in nanometers) ratios of 0.77:1.43 and 0.90:1.60, respectively. So in this paper we mainly studied structural, electronic and optical properties of a (7,7)@(12,12) DWNT bundle (with a inner-to-outer diameter ratio of 0.949:1.627) under hydrostatic pressure by combining a classical force field, tight-binding (TB) model and first-principles method. The paper is organized as follows: In Sec. II we introduce the numerical methods, the results and the discussion are given in Sec. III, and some concluding remarks are given in Sec. IV.

# II. Methods

First we studied the changes of the geometrical structure of DWNT bundles induced by a combination of hydrostatic pressure and van der Waals forces. The zero-temperature structural minimizations of the enthalpy ($H = U + PV$) were carried out on a supercell containing $2 \times 2 \times 2$ DWNTs using a universal force field (UFF) method [17, 18]. In order to induce the SPT, a step-wise increasing hydrostatic pressure was applied to the DWNT bundle and the enthalpy of the DWNT bundle was minimized after each pressure increment. Then density-functional theory (DFT) calculations in the framework of the local density approximation (LDA) were used to investigate electronic and optical properties. We used the projector augmented wave (PAW) method [19, 20] with a uniform energy cutoff of 500 eV. The maximum spacing between $k$ points was 0.03 Å$^{-1}$ and the smearing width was taken to be 0.04 eV in the ground state.

For the (7,7)@(12,12) DWNT bundle, we have performed structural optimizations at zero external pressure using both the UFF method and the first principles method. We obtained a very good agreement (see Table I), indicating that the UFF method is well suited for the system under consideration.

In addition to DFT-LDA we also employ TB calculations to study the electronic structure of CNTs. The basis set consists of one $s$ and three $p$ orbitals per atom, allowing to take $\sigma - \pi$ electron hybridization into account [21–23]. The on-site energies are $\epsilon_s^0 = -7.3$ eV and $\epsilon_p^0 = 0.0$ eV for the $s$ and $p$ states, respectively. The Slater-Koster hopping parameters for the first nearest-neighbors are taken as $V_{ss\sigma} = -4.30$ eV, $V_{sp\sigma} = 4.98$ eV, $V_{pp\sigma} = 6.38$ eV and $V_{pp\pi} = -2.66$ eV. The interactions with higher order nearest-neighbors are taken into account by using $V_{ss\sigma} = -0.18Y$ eV, $V_{sp\sigma} = 0.21Y$ eV, $V_{pp\sigma} = 0.27Y$ eV and $V_{pp\pi} = -0.11Y$ eV, where $Y = (3.335/r_{ij})^2$ is a scaling factor, depending on the interatomic distance $r_{ij}$ (in unit of Å). In our calculation, the cutoff of $r_{ij} = 3.9$ Å is used in order to include Van der Waals forces.

### III. Results and Discussion

In Fig. 1 the volume change of the (7,7)@(12,12) DWNT bundle was determined as function of the applied hydrostatic pressure using the UFF method. At a critical pressure of 10.6 GPa a discontinuous SPT can be seen. At that point the DWNTs spontaneously collapse into ones with a peanut-shaped cross sections, where the separation between two opposite parallel walls of the inner (7,7) tubes is approximately equal to the distance between layers in turbostratic graphite (3.4 Å). We call this critical pressure the *collapse pressure* $P_d$, and the corresponding structure the *collapsed structure*. Our simulations further indicate that (7,7) and (12,12) SWNT bundles [shown in Fig. 1] collapse at $P_d = 7$ and 2.4 GPa, respectively.

The collapse pressure of the (7,7)@(12,12) DWNT bundle is higher than the ones of (7,7) and (12,12) SWNT bundles. Comparing the transition pressure of the inner tube with that of the corresponding DWNT yields a ratio of 0.66. This ratio obviously relates to the radius of the inner and outer tubes, *e.g.* it is 0.75 and 0.90 for (11,11)@(16,16) and (15,15)@(20,20) DWNT bundles, respectively. This means that the outer tube acts as a "protection shield", and the inner tube supports the outer one and increases its structural stability. This picture is also consistent with the experimental results [2–4]. However, due to the decreasing radial stiffness of SWNTs, this effect will become weak as the tube radius increases. Furthermore, our simulations (see Fig. 1) reveal differences in the tubular cross-sections shortly before the collapse: the cross-sections of the (7,7)@(12,12) DWNT bundle and the (12,12) SWNT bundle are hexagon-like, with almost equal sides and corners, however the cross-section of the (7,7) SWNT bundle is elliptic. The reason is that the (12,12) tube has a $C_6$ rotational axis, which can be matched with the hexagonal lattice symmetry. Therefore, the symmetry of the outer tube is an important factor, deciding the cross sectional shape of the DWNT under pressure. On the other hand, the effect of their inner tube is very small.

So far, three possible superstructures of collapsed CNTs have been studied by theory; they are called *parallel* [13], *herringbone* [12], and *in-between* [24]. For SWNT bundles, previous results of us [24] indicate that the parallel and the in-between collapsed structures are more favorable in enthalpy than the herringbone one. But which of those configurations is most favorable in DWNT bundles? Figure 2(a)-(c) shows the three different collapsed superstructures of the (7,7)@(12,12) DWNT bundle after the SPT at 11 GPa. It must be noted that for the parallel structure all nanotubes are actually equivalent (see Figs. 2(a) and 2(d)). However the supercells of the other two structures contain two inequivalent tubes. The results of our simulation on collapsed DWNTs superstructures are listed in Table II. We find that the in-between and parallel structures are almost degenerate in enthalpy, having a difference of only 0.1 meV per atom. Both are more favorable than the herringbone structure concerning the total energy, the PV term, and the enthalpy. These results are in qualitative agreement with those found for SWNT bundles, but now the largest difference in enthalpy among the three structures is about 9.8 meV per atom, which is almost twice of that found for SWNT bundles [24], indicating that the herringbone structure in DWNT bundle is even more unstable than in the case of a SWNT bundle. Comparing the parallel and in-between structures of the (7,7)@(12,12) DWNT bundle in Fig. 2(a) and 2(b) we find that the structural differences are very small. Further simulations revealed that these differences become bigger as the nanotube radius increases. This can be seen in Figs. 2(d)-2(f), where we show the three collapsed structures of a (11,11)@(16,16) DWNT bundle at 3.5 GPa after SPT; the simulation results are also given in Table II.

The electronic band-structures of individual (7,7) and (12,12) SWNTs, and the (7,7)@(12,12) DWNT bundle at 0, 9, and 11 GPa, calculated with DFT-LDA are given in Figs. 3(a)-3(e) [Fig. 3(e) corresponds to the parallel collapsed structure], along symmetry lines in the Brillouin zone (BZ). The selected high symmetry $k$-points are $\Gamma = (0,0,0)$, $X = (0,0,0.5)$, $B = (0.5, 0.0, 0.0)$ and $Q = (0, 0.5, 0.5)$. The intertube van der Waals forces and the structural deformations break the rotational symmetry of the SWNTs even at zero pressure. This splits up the $k_z$ energy bands of DWNTs near the Fermi Level. Figure 3(c) shows that these bands are different from a simple superposition of two individual SWNTs' energy bands at zero pressure [compare with Figs. 3(a) and 3(b)]. The band differences near the Fermi level are small at 0 and 9 GPa before the SPT, but at 11 GPa, after the SPT, a pseudogap appears along the $\Gamma X$ direction.

In a bundle, the interactions between nanotubes do not only alter the band structure along $\Gamma X$, but also cause a dispersion in the perpendicular direction. The theoretical work of Reich *et al.* [25] indicates that a perpendicular dispersion is expected to broaden the electronic density of states and the optical absorption bands in nanotube bundles. It is interesting to study the changes of the perpendicular dispersion upon applying a hydrostatic pressure on a DWNT bundle. In Fig. 3(c)-3(e), we can see that with the increasing pressure the perpendicular dispersion along $B\Gamma$ and $XQ$ comes closer to the Fermi Level. The right-hand panels of these subfigures show the perpendicular dispersion along $PU$ close to the Fermi Level at $k_z = 0.317$ [$\Delta = (0,0,0.317)$] at 0 and 9 GPa, and at $k_z = 0.3$ and $0.329$ [$\Delta_1 = (0,0,0.3)$ and $\Delta_2 = (0,0,0.329)$] at 11 GPa. The $k$-points $P$ and $U$ are located at $(0.5, 0.0, k_z)$ and $(0, 0.5, k_z)$, respectively. Applying pressure obviously steepens the dispersion perpendicular to the $k_z$ direction, leads to an increase of the bandwidths, and lifts the degeneracy of some bands at 9 GPa before the SPT. After the SPT the structural symmetry is lowered even further and more and more bands appear in the energy range $-2.0\ldots 2.0$ eV. However the bandwidths are smaller after the SPT than before.

The electronic band-structure along the $\Gamma X$ of the (7,7)@(12,12) DWNT bundle at 0, 9, and 11 GPa (the latter pressure again corresponds to the parallel collapsed structure), are also calculated by the TB method and are given in Figs. 4(a)-4(c), respectively. We find that the TB results qualitatively agree with the first principles calculations. TB calculations for the deformed inner (7,7) and outer (12,12) tubes those are separated each other from corresponding DWNT, revealed that energy bands change slightly from 0 to 9 GPa and that at 11 GPa an energy gap of 0.08 eV opens at the Fermi level of inner (7,7) tube (along the $\Gamma X$ direction) while the outer one has no gap.

In order to study the influence of the nanotube diameter on the structural changes and the electronic properties under hydrostatic pressure, we also simulated a (15,15)@(20,20) DWNT using the UFF method. The volume changes as function of the applied hydrostatic pressure are given in Fig. 5. We can see that the pressure-response of the (15,15)@(20,20) DWNT bundle is more complex. The bundle undergoes two different types of SPTs at transition pressures of 1.25 and 1.44 GPa, respectively. These SPTs are accompanied by cross-sectional changes from a deformed hexagon to a "racetrack" shape at 1.25 GPa, going over to a "peanut" shape and collapsing completely at 1.44 GPa. Furthermore, one notices slight first-order discontinuities in the volume change among the different racetrack-shaped cross sections. Our simulations indicate that the (15,15) and (20,20) SWNT bundles [shown in Fig. 5] collapse at $P_d = 1.3$ and 0.54 GPa, respectively. However, before the (20,20) SWNT bundle collapses it undergoes a first SPT at 0.2 GPa which is accompanied by a change of the cross-section from a deformed hexagon to a racetrack shape. The (15,15) tube can keep the hexagonal shape before the SPT and, more importantly, it undergoes only one SPT to reach the stable collapsed structure. Those results are similar to the ones found in the (12,12) SWNT system. One reason is the $C_3$ symmetry of the (15,15) tube, which partly matches the hexagonal lattice symmetry. Finally, we investigated the electronic band structures of the (15,15)@(20,20) DWNT bundle along the $\Gamma X$ direction at 0, 1, 1.35, and 1.6 GPa using the TB method. The results are shown in Fig. 6. We find that a pseudogap appears both at 1.35 GPa (after the first SPT) and at 1.6 GPa (after the second SPT). After the collapse (at 1.6 GPa) the inner (15,15) tube becomes semiconducting with an energy gap of 0.11 eV. This behavior is analogous to that of the collapsed (7,7)@(12,12) DWNT bundle.

Changes of the electronic structure are well reflected in optical absorption spectra. We calculated these spectra for the individual (7,7) and (12,12) SWNTs ($k$-point sampling: $1 \times 1 \times 120$) and for the DWNT bundles ($k$-point sampling: 0.02 Å$^{-1}$); the results are shown in Figs. 7(a)-7(e). For the individual (7,7) and (12,12) SWNTs [Fig. 7(a) and 7(b)] the characteristic absorption peaks polarized along the tube direction (Z direction) are centered at 2.1284, 3.366 and 3.7876 eV and at 1.3532, 2.448, 3.2096, 3.706 and 3.944 eV, respectively. These peaks can also be identified in the absorption spectra of the corresponding (7,7)@(12,12) DWNT bundle [see Fig. 7(c)], but they are slightly shifted because of the band splitting induced by the tube-tube interaction in the bundle (see above). Additionally, strong optical absorptions emerge in the energy region 0...0.5 eV due to the presence of pseudogaps along directions parallel to $\Gamma X$. This effect was predicted by Paul Delaney *et al.* [26] and was confirmed by experiment [27] later. With the increasing pressure the cross section of the (7,7)@(12,12) DWNT bundle becomes hexagonal at 9 GPa and peanut-like shaped at 11 GPa, which further lowers the symmetry of the structure. This leads to a significant broadening of the characteristic absorption peaks at 9 and 11 GPa [see Figs. 7(d)-7(e)]. The first absorption peak of the outer tube, which is at 1.1764 eV at zero pressure [Fig. 7(c)], is slightly red-shifted to 1.0404 eV at 9 GPa and remains sharp [Fig. 7(d)]. But at 11 GPa, just after the SPT, it abruptly becomes broader and is blue-shifted to 1.2444 eV [see Fig. 7(e)]. Thus the behavior of the first characteristic absorption peak reflects the SPTs of the (7,7)@(12,12) DWNT bundle and can be considered as a fingerprint.

Furthermore, Figs. 7(c)-7(e) show that there is a large difference in the optical absorption anisotropy for the (7,7)@(12,12) DWNT bundle at 0, 9, and 11 GPa. This is similar to our previous result on SWNT bundles [24]. The optical responses polarized along X and Y directions are almost the same in the (7,7)@(12,12) DWNT bundle at 0 and 9 GPa (before the SPT) but differ significantly in the parallel collapsed structure at 11 GPa due to the changes of structure and symmetry. For the collapsed structure the optical response in the low-energy region is weaker along the X direction than along Y. To understand that let us consider that the flat segment of the collapsed cross section [see Fig. 2(a)] is very similar to a graphite sheet. It is well known that in graphite, the optical absorption component polarized perpendicular to the carbon-layer is much smaller than that parallel to the layer [see Ref. 28]. In Fig. 2(a) it can be seen that the Y axis is more parallel to the flattened part of the DWNT than the X axis. Therefore the Y component of the optical absorption is stronger than the X component [see Fig. 7(e)]. We have done the same calculations for other DWNT

bundles with bigger radii [(9,9)@(14,14), (10,10)@(15,15), (11,11)@(16,16), (12,12)@(17,17), and (13,13)@(18,18)] where the flattened segments of the collapsed DWNT are bigger, and it is found that the absorption components that relate to the flattened segment also increase in magnitude. This indicates that the graphite-like characteristic anisotropy becomes more and more noticeable as the radii of the collapsed DWNT bundles increase.

## IV. Summary

We have studied structural, electronic, and optical properties of a (7,7)@(12,12) DWNT bundle under hydrostatic pressure, by using a classical force field approach, a TB model and first-principle calculations. We have found that the outer tube acts as a "protection shield" for the inner tube, and that the inner tube increases the structural stability of the outer tube, i.e., the ability to resist an external pressure. As in SWNT bundles, the "parallel" and "in-between" collapsed superstructures of DWNT bundles are more stable than the "herringbone" one. A SPT induces a pseudogap in the electronic band structure along the $\Gamma X$ direction in reciprocal space. The optical properties of the DWNT bundle change significantly after the collapse and a strong optical anisotropy appears, which can be considered as a fingerprint of the SPT. Furthermore, UFF simulations of DWNT bundles reveal that, depending on the symmetry of the outer tube, two SPTs can exist.

Acknowledgments: This work was supported by the Natural Science Foundation of China under Grant No. 10474035 and 90503012, and also by the State Key program of China through Grant No. 2004CB619004.

[1] Liu Yang and Jie Han, *Phys. Rev. Lett.*, **85** (2000) 154.

[2] P. Puech, H. Hubel, D.J. Dunstan, R.R. Bacsa, C. Laurent, and W.S. Bacsa, *Phys. Rev. Lett.*, **93** (2004) 095506.

[3] J. Arvanitidis, D. Christofilos, K. Papagelis, K.S. Andrikopoulos, T. Takenobu, Y. Iwasa, H. Kataura, S. Ves, and G.A. Kourouklis, *Phys. Rev. B*, **71** (2005) 125404.

[4] J. Arvanitidis, D. Christofilos, K. Papagelis, T. Takenobu, Y. Iwasa, H. Kataura, S. Ves, and G.A. Kourouklis, *Phys. Rev. B*, **72**, 193411 (2005).

[5] U.D. Venkateswaran, A.M. Rao, E. Richter, M. Menon, A. Rinzler, R.E. Smalley, and P.C. Eklund, *Phys. Rev. B*, **59** (1999) 10928.

[6] M.J. Peters, L.E. McNeil, J.P. Lu, and D. Kahn, *Phys. Rev. B*, **61** (2000) 5939.

[7] C. Thomsen, S. Reich, A.R. Goni, H. Jantoljak, P.M. Rafailov, I. Loa, K. Syassen, C. Journet, P. Bernier, *Phys. Status Solidi B*, **215** (1999) 435.

[8] C. Thomsen, S. Reich, H. Jantoljak, I. Loa, K. Syassen, M. Burghard, G.S. Duesberg, and S. Roth, *Appl. Phys. A* **69** (1999) 309.

[9] J. Tang, L.-C. Qin, T. Sasaki, M. Yudasaka, A. Matsushita, and S. Iijima, *Phys. Rev. Lett.*, **85** (2000) 1887.

[10] S. Kazaoui, N. Minami, H. Yamawaki, K. Aoki, H. Kataura, and Y. Achiba, *Phys. Rev. B*, **62** (2000) 1643.

[11] M.H.F. Sluiter, V. Kumar, and Y. Kawazoe, *Phys. Rev. B*, **65** (2002) 161402; M.H.F. Sluiter, and Y. Kawazoe, *ibid* **69** (2004) 224111.

[12] J.A. Elliott, J.K.W. Sandler, A.H. Windle, R.J. Young, and M.S.P. Shaffer, *Phys. Rev. Lett.*, **92** (2004) 095501.

[13] X.H. Zhang, Z.F. Liu, and X.G. Gong, *Phys. Rev. Lett.*, **93** (2004) 149601; Siu-Pang Chan, Wai-Leung Yim, X.G. Gong, and Zhi-Feng Liu, *Phys. Rev. B*, **68** (2003) 075404; X.H. Zhang, D.Y. Sun, Z.F. Liu, and X.G. Gong, *ibid* **70** (2004) 035422; X. Ye, D.Y. Sun, and X.G. Gong, *ibid* **72** (2005) 035454.

[14] V. Gadagkar, P.K. Maiti, Y. Lansac, A. Jagota and A.K. Sood, *Phys. Rev. B*, **73** (2006) 085402.

[15] Xiaoping Yang, Gang Wu, and Jinming Dong, *Appl. Phys. Lett.*, **89** (2006) 113101.

[16] M. Endo, H. Muramatsu, T. Hayashi, Y.A. Kim, M. Terrones, M.S. Dresselhaus, *Nature*, **433** (2004) 476.

[17] A.K. Rappe, C.J. Casewit, K.S. Colwell, W.A. Goddard, W.M. Skiff, *J. Am. Chem. Soc.*, **114** (1992) 10024.

[18] N. Yao, V. Lordi, *J. Appl. Phys.*, **84** (1998) 1939.

[19] G. Kresse, J. Furhmuller, Software VASP, Vienna (1999); G. Kresse, J. Hafner, *Phys. Rev. B*, **47** (1993) R558; *ibid* **49** (1994) 14251; *ibid* **54** (1996) 11169; *Comput. Mat. Science*, **6** (1996) 15.

[20] P.E. Blöchl, *Phys. Rev. B*, **50** (1994) 17953; G. Kresse and D. Joubert, *ibid* **59** (1999) 1758.

[21] J.C. Charlier, Ph. Lambin, and T. W. Ebbesen, *Phys. Rev. B*, **54** (1996) 8377.

[22] H.S. Sim, C.J. Park, and K. J. Chang, *Phys. Rev. B*, **63** (2001) 073402.

[23] X. Blase, L.X. Benedict, E.L. Shirley, and S. G. Louie, *Phys. Rev. Lett.*, **72** (1994) 1878.

[24] Xiaoping Yang, Gang Wu, Jian Zhou, and Jinming Dong, *Phys. Rev. B*, **73** (2006) 235403.

[25] S. Reich, C. Thomsen, and P. Ordejón, *Phys. Rev. B*, **65** (2002) 155411.

[26] Paul Delaney, Hyoung Joon Choi, Jisoon Ihm, Steven G. Louie and Marvin L. Cohen, Nature **391** (1998) 466; *Phys. Rev. B*, **60** (1999) 7899.

[27] Min Ouyang, Jin-Lin Huang, Chin Li Cheung, Charles M. Lieber, *Science*, **292** (2001) 702.

[28] G.Y. Guo, K.C. Chu, Ding-sheng Wang, Chun-gang Duan, *Phys. Rev. B*, **69** (1996) 205416.

# TABLE

TABLE I: Calculated equilibrium lattice parameters of the (7,7)@(12,12) DWNT bundle at zero pressure obtained by UFF and DFT-LDA calculations. a, b and c are the lattice constants of the DWNT bundles, and $\alpha$, $\beta$ and $\gamma$ are the angles between two lattice vectors.

|     | a     | b     | c    | $\alpha$ | $\beta$ | $\gamma$ |
|-----|-------|-------|------|----------|---------|----------|
| UFF | 19.26 | 19.26 | 2.44 | 90.00    | 90.00   | 119.98   |
| LDA | 19.26 | 19.26 | 2.45 | 90.00    | 90.00   | 119.98   |

TABLE II: Total energies (E) and enthalpies (H) of the "parallel", the "in-between", and the "herringbone" structures at the collapse pressure and at zero temperature.

| DWNT bundle | E/atom (eV) | PV/atom (eV) | H/atom (eV) |
|---|---|---|---|
| (7,7)@(12,12)-Parallel | 0.2814 | 0.5728 | 0.8541 |
| (7,7)@(12,12)-In-between | 0.2809 | 0.5730 | 0.8540 |
| (7,7)@(12,12)-Herringbone | 0.2862 | 0.5777 | 0.8639 |
| (11,11)@(16,16)-Parallel | 0.2078 | 0.1926 | 0.4004 |
| (11,11)@(16,16)-In-between | 0.2078 | 0.1925 | 0.4003 |
| (11,11)@(16,16)-Herringbone | 0.2117 | 0.1983 | 0.4100 |

**Figure Caption**

Fig. 1 (Color online) The Loading curves of the (7,7)@(12,12) DWNT bundle and the corresponding SWNT bundles as a function of hydrostatic pressure.

Fig. 2 (Color online) Snapshots of the cross-sections of DWNT bundles after the collapse: (a)(d) the "parallel" structure, (b)(e) the "in-between" structure, and (c)(f) the "herringbone" structure. The boxes represent the unit cells. Inequivalent nanotubes are colored differently in panel (b).

Fig. 3 (Color online) The electronic band structures along symmetry lines obtained by first-principles (DFT-LDA) calculations. (a) An individual (7,7) SWNT at 0 GPa, (b) an individual (12,12) SWNT at 0 GPa, (c) (7,7)@(12,12) DWNT bundle at 0 GPa, (d) (7,7)@(12,12) DWNT bundle at 9 GPa, and (e) (7,7)@(12,12) DWNT bundle at 11 GPa. The Fermi level is set equal to zero.

Fig. 4 (Color online) The electronic band structures of the (7,7)@(12,12) DWNT bundle obtained by TB calculations. (a) At 0 GPa, (b) at 9 GPa, (c) at 11 GPa. The Fermi level is set equal to zero.

Fig. 5 (Color online) The loading curves for the (15,15)@(20,20) DWNT bundle and the corresponding SWNT bundles as a function of hydrostatic pressure.

Fig. 6 (Color online) The electronic band structures of the (15,15)@(20,20) DWNT bundle obtained by TB calculations. (a) At 0 GPa, (b) at 1 GPa, (c) at 1.35 GPa, and (d) at 1.6 GPa. The Fermi level is set equal to zero.

Fig. 7 (Color online) The imaginary part (absorptive part) of the dielectric function polarized along the X, Y and Z directions obtained by first-principles (DFT-LDA) calculations. (a) An individual (7,7) SWNT at 0 GPa, (b) an individual (12,12) SWNT at 0 GPa, (c) (7,7)@(12,12) DWNT bundle at 0 GPa, (d) (7,7)@(12,12) DWNT bundle at 9 GPa, (e) (7,7)@(12,12) DWNT bundle at 11 GPa.

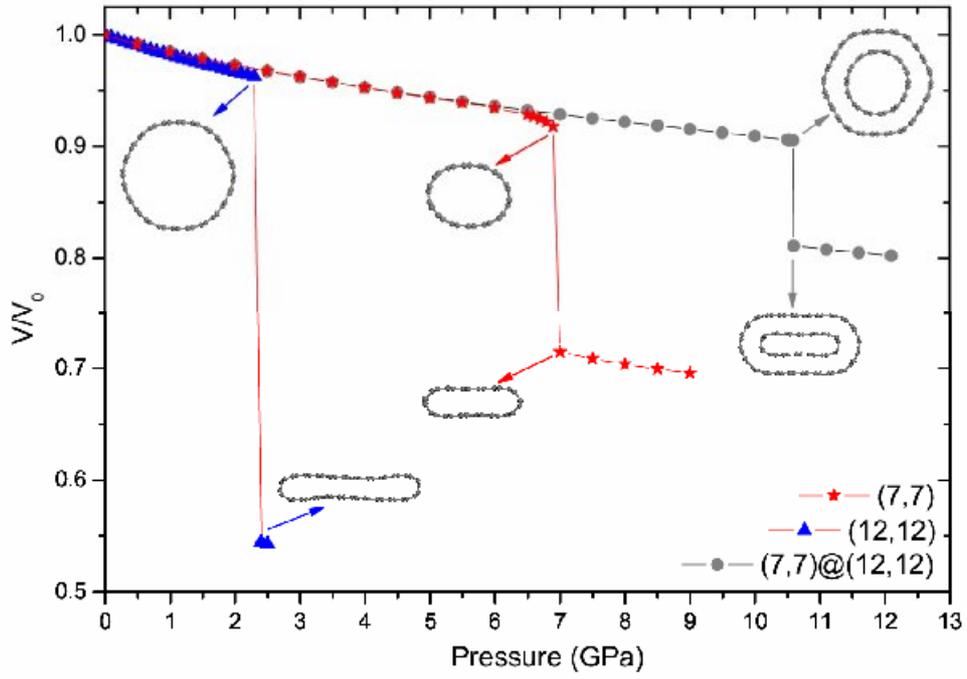

**Fig. 1**

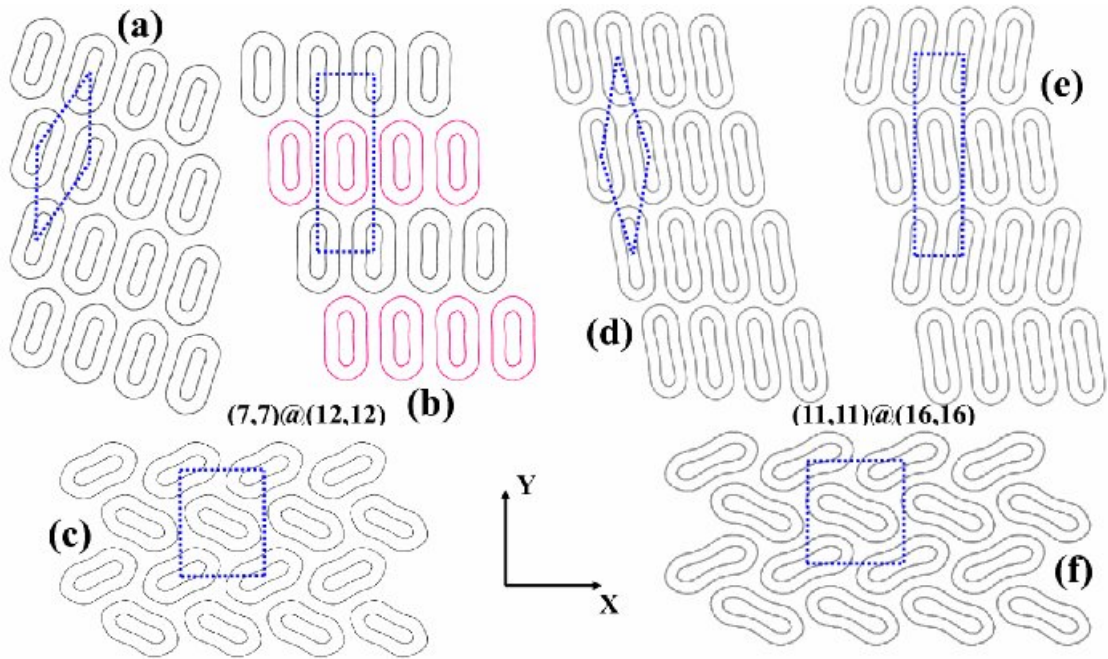

**Fig. 2**

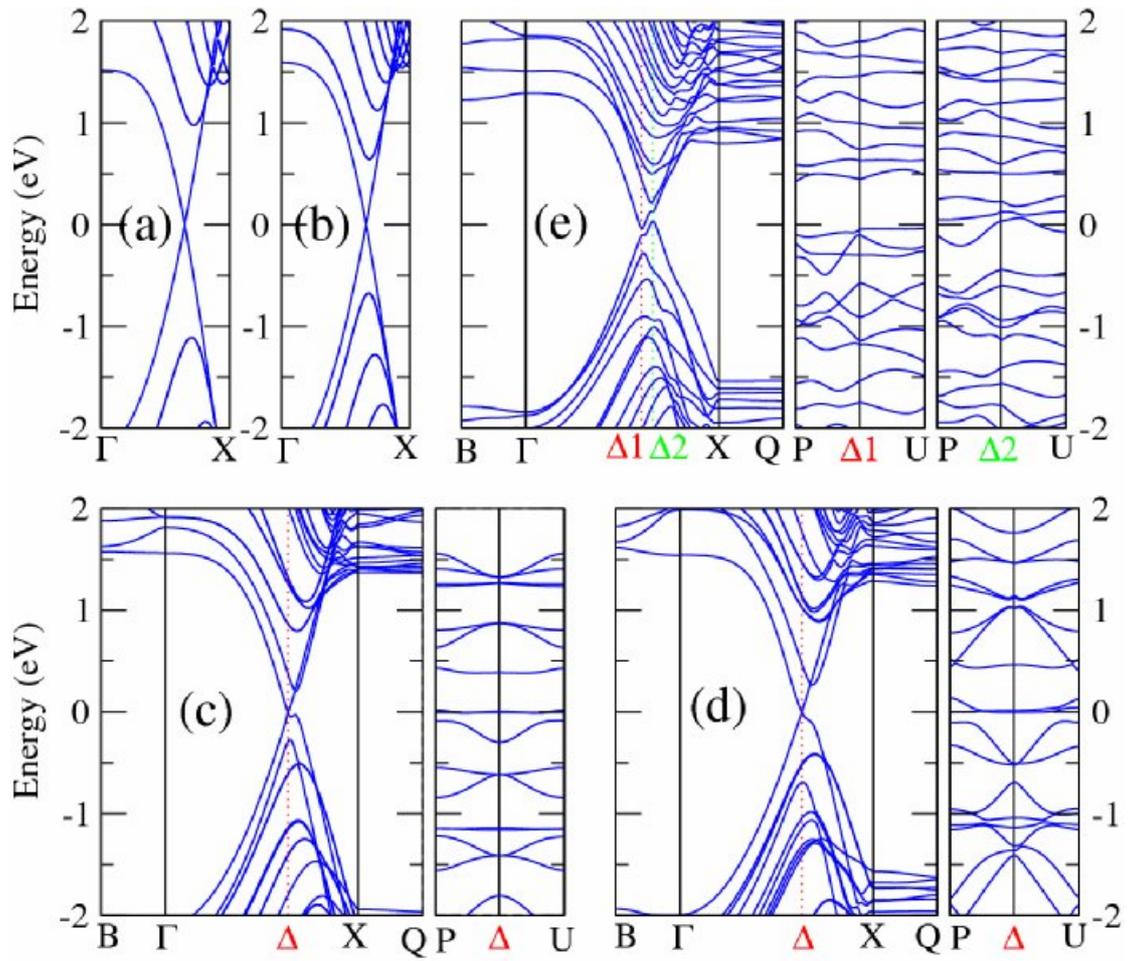

**Fig. 3**

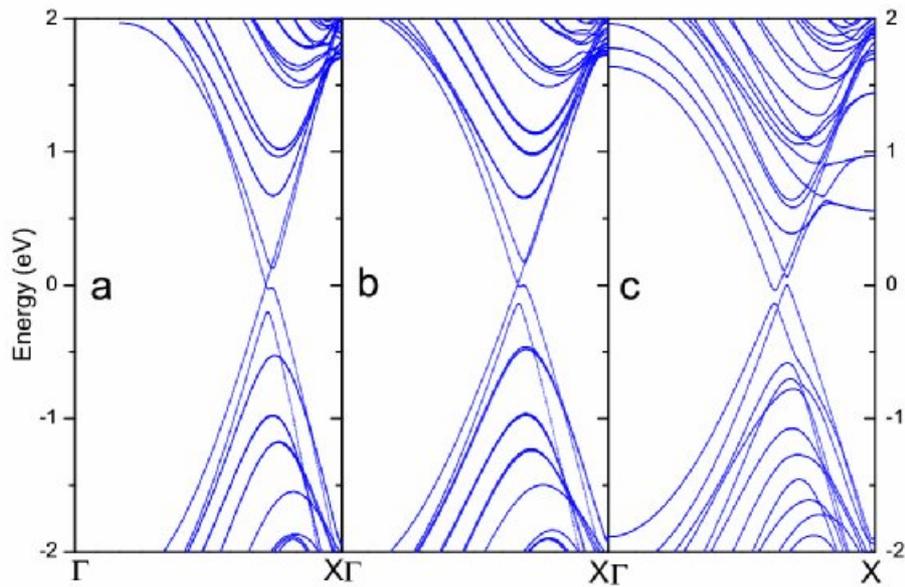

**Fig. 4**

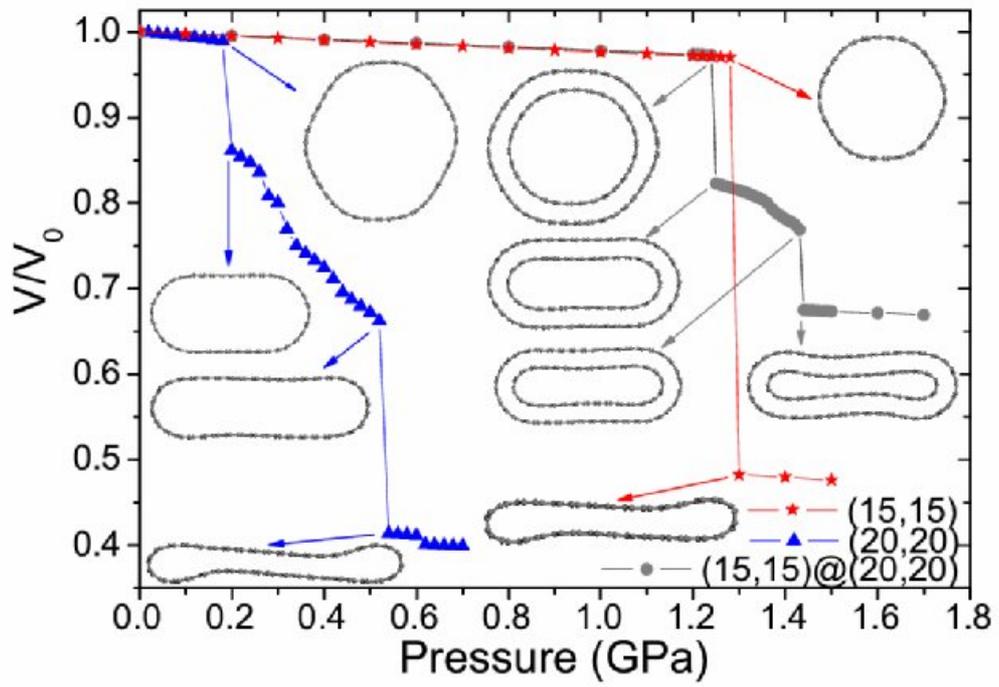

**Fig. 5**

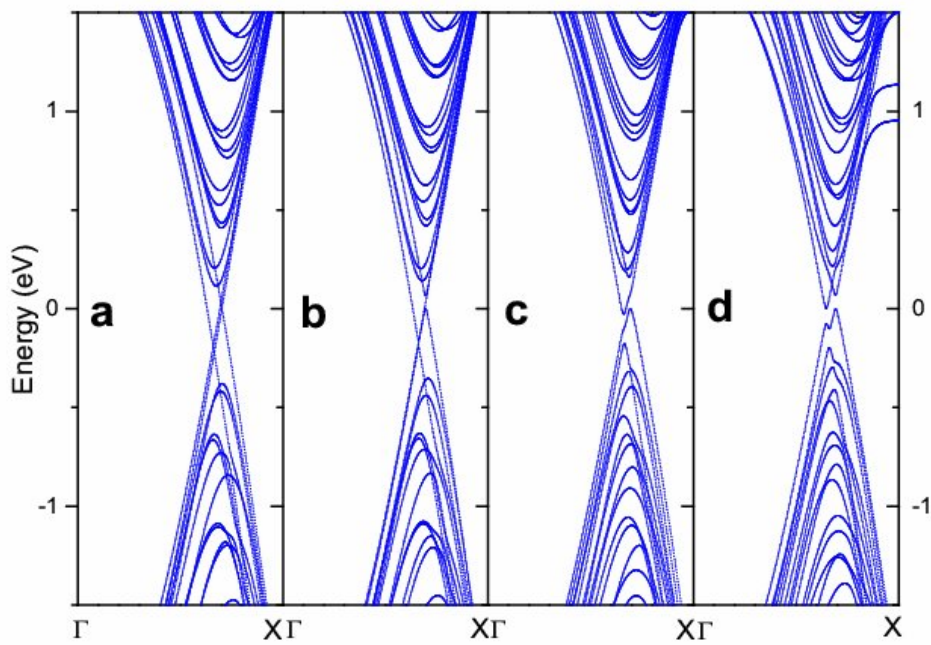

**Fig. 6**

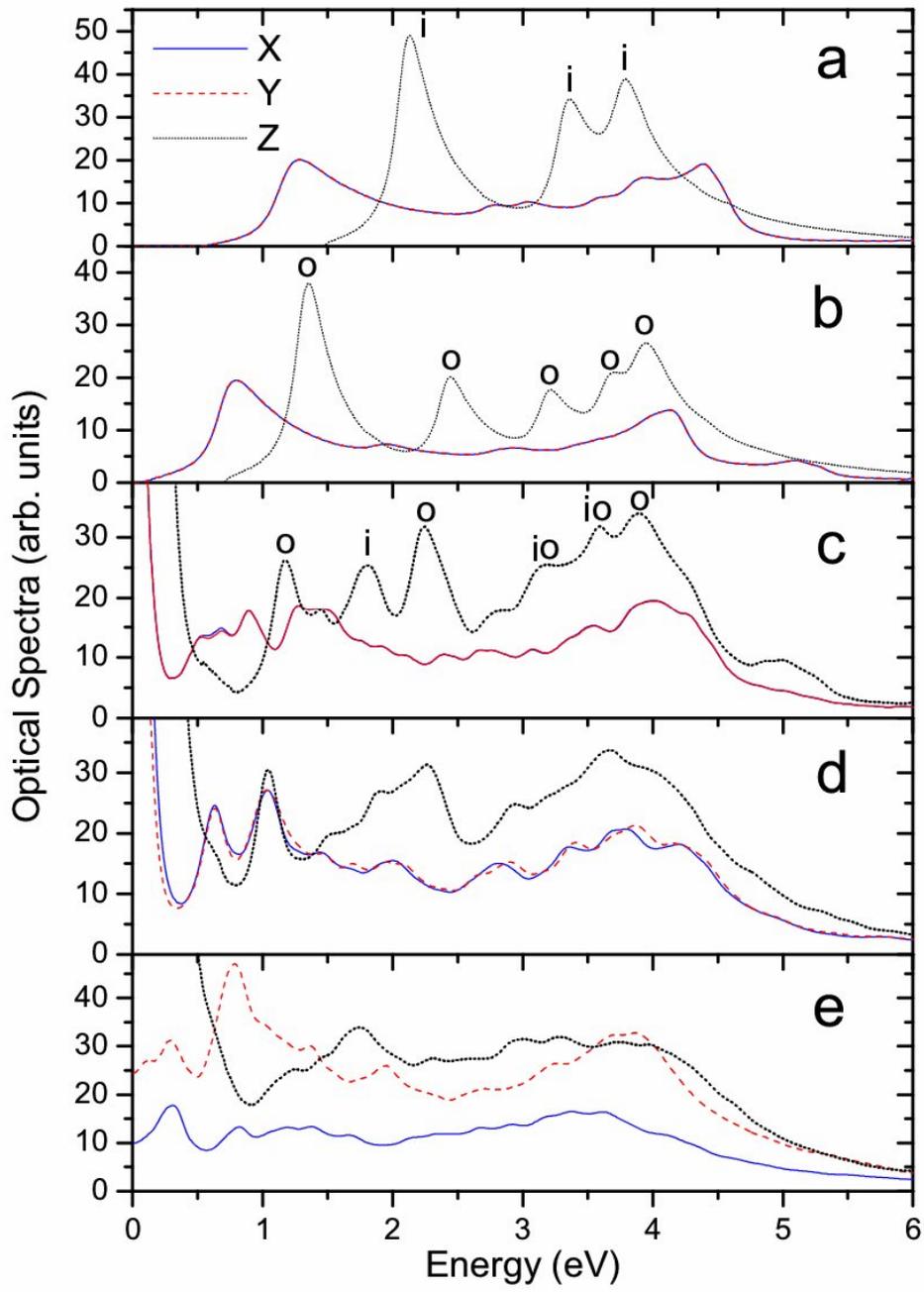

**Fig. 7**